\documentclass[12pt,epsf,amssymb]{article}
\usepackage{epsfig,psfrag}
\usepackage{graphicx}
\usepackage{amsfonts}
\usepackage{amsmath}
\usepackage[usenames]{color}
\usepackage{amssymb}

\begin{document}
\baselineskip 0.6cm
\newcommand{\gsim}{ \mathop{}_{\textstyle \sim}^{\textstyle >} }
\newcommand{\lsim}{ \mathop{}_{\textstyle \sim}^{\textstyle 3<} }
\newcommand{\vev}[1]{ \left\langle {#1} \right\rangle }
\newcommand{\bra}[1]{ \langle {#1} | }
\newcommand{\ket}[1]{ | {#1} \rangle }
\newcommand{\Dsl}{\mbox{\ooalign{\hfil/\hfil\crcr$D$}}}
\newcommand{\nequiv}{\mbox{\ooalign{\hfil/\hfil\crcr$\equiv$}}}
\newcommand{\nsupset}{\mbox{\ooalign{\hfil/\hfil\crcr$\supset$}}}
\newcommand{\nni}{\mbox{\ooalign{\hfil/\hfil\crcr$\ni$}}}
\newcommand{\EV}{ {\rm eV} }
\newcommand{\KEV}{ {\rm keV} }
\newcommand{\MEV}{ {\rm MeV} }
\newcommand{\GEV}{ {\rm GeV} }
\newcommand{\TEV}{ {\rm TeV} }

\def\diag{\mathop{\rm diag}\nolimits}
\def\tr{\mathop{\rm tr}}

\def\Spin{\mathop{\rm Spin}}
\def\SO{\mathop{\rm SO}}
\def\O{\mathop{\rm O}}
\def\SU{\mathop{\rm SU}}
\def\U{\mathop{\rm U}}
\def\Sp{\mathop{\rm Sp}}
\def\SL{\mathop{\rm SL}}

\def\change#1#2{{\color{blue}#1}{\color{red} #2}\color{black}\hbox{}}


\begin{titlepage}

\begin{flushright}
LTH/775
\end{flushright}

\vskip 2cm
\begin{center}
{\large \bf  SQCD Vacua and Geometrical Engineering}
\vskip 1.2cm
Radu Tatar and Ben Wetenhall

\vskip 0.4cm

{\it Division of Theoretical Physics, Department of Mathematical Sciences

The University of Liverpool,
Liverpool,~L69 3BX, England, U.K.

rtatar@liverpool.ac.uk,~~benweten@liv.ac.uk}

\vskip 1.5cm

\abstract{We consider the geometrical engineering constructions for the N = 1 SQCD vacua recently proposed by Giveon and Kutasov.
After one T-duality, the geometries with wrapped D5 branes become N = 1 brane configurations with NS branes and D4 branes. 
The field theories encoded by the geometries contain extra
massive adjoint fields for the  flavor group. After performing a flop, the geometries contain 
branes, antibranes and branes wrapped on non-holomorphic cycles.
The various tachyon condensations between pairs of wrapped D5 branes and anti D5 branes together with deformations of the cycles give rise to a variety of 
supersymmetric and metastable non-supersymmetric vacua. 
} 

\end{center}
\end{titlepage}


\section{Introduction}

Over the last year many directions have been investigated  to describe metastable vacua in  ${\cal N} =1$ SUSY theories after the 
work of \cite{iss}. String theory appears to be a useful tool to obtain a large number of such solutions.
String theory constructions contain D4 branes, D6 branes and NS branes as it was well-known for the
SUSY vacua \cite{giku}. Alternative constructions involve wrapped D5 branes on $P^1$ cycles inside resolved Calabi-Yau 
manifolds. These two pictures are related by T-duality along the radial direction of the various $P^1$ cycles, as 
discussed extensively in \cite{dot1}-\cite{dot2}.

The situation becomes less clear when considering the metastable vacua. In terms of the corresponding geometry (or the 
T-dual brane configuration), there are two different types of metastable vacua obtained from string theory:

1) wrap D5 branes and anti D5-branes on the same $P^1$ cycle of the Calabi-Yau manifolds. The corresponding T-dual was first mentioned in 
\cite{tw1} and then clarified in \cite{marsano}. These geometries have the usual geometric transitions and 
the strongly coupled theories are described in terms of deformed geometries with positive and negative fluxes.
Their T-dual configurations contain D4 branes and anti D4 branes.

The case of  D5 branes and anti D5-branes on the same $P^1$ cycle has also been nicely extended to the case when the 
 D5 branes and anti D5-branes are wrapped on different $P^1$ cycles which are on the same homology class. \cite{agava1}.

2) wrap some D5 branes on the  $P^1$ cycles of the Calabi-Yau manifolds and some other D5 branes on non-holomorphic deformations of 
the  $P^1$ cycles of the Calabi-Yau manifolds. \footnote{Some possible embedding of these non-holomorphic cycles 
has been discussed in \cite{oz1} in the context of M-theory lifting of type IIA configurations.} The T-dual of this geometry 
are the brane configurations for the ISS type models discussed in \cite{oo2}. Other models for
field theories with  metastable vacua were built in \cite{ahn}.

Very recently, the authors of \cite{giku1,giku2} have extended the results of \cite{iss} to the case when there is an extra 
massive adjoint field in the flavor group. This adjoint field is integrated out in the electric theory and is identified with the 
gauge singlet in the magnetic theory. Both sets of SUSY and metastable non SUSY vacua are enlarged by introducing this extra adjoint field.
The results of \cite{giku1,giku2} were obtained by using brane configurations with NS branes, D4 branes and
D6 branes. 

In the present work we are going to give an alternative IIB geometrical engineering picture together with its T-dual IIA 
configuration with only NS branes and D4 branes. We obtain the  ${\cal N} = 1$ theories as deformations of 
 ${\cal N} = 2$ theories. By starting with an
  ${\cal N} =2, SU(N_f) \times SU(N_c)$ theory, the breaking of SUSY in \cite{iss} is obtained for
zero vevs of electric quarks  
\begin{equation}
{\cal N} =2, SU(N_f) \times SU(N_c) \rightarrow {\cal N} =1, SU(N_f) \times SU(N_c)
\end{equation}
whereas the breaking in \cite{giku1,giku2} allows vevs for the electric quarks
\begin{equation}
{\cal N} =2, SU(N_f) \times SU(N_c) \rightarrow {\cal N} =1, SU(N_f-k) \times SU(N_c-k) \times SU(k),~~k~\ne~0
\end{equation}
for a general value of $k < N_f~k<N_c$. 
This can be easily seen from the brane configurations with NS branes and D4 branes. If the masses for the adjoint fields 
of $SU(N_f)$ and $SU(N_c)$ are infinite, then the vacuum expectation values for $k$ bifundamental fields
\begin{equation}
<Q \tilde{Q}> = \frac{\mu}{\frac{1}{m_{N_f}} +  \frac{1}{m_{N_c}}}
\end{equation}
are infinite and the $ SU(N_f-k) \times SU(N_c-k)$ are completely decoupled from the $SU(k)$. 

The Seiberg duality is a flop in the geometry and the dynamics of branes is related to a 
tachyon condensation between D5 branes and anti D5 branes \cite{tw2}. In the present discussion the tachyon condensation 
explains the ranks of the groups in the magnetic theory but also determine a reorientation of the cycle such they 
remain holomorphic in the magnetic theory. This is related to the possibility that the cycles slide along their 
normal bundles, which was not allowed for the fixed cycles of the geometries for the \cite{iss} model. 
The metastable solutions of \cite{giku1,giku2} when there are some extra D5 branes wrapped on certain 2-cycles. In the Seiberg dual picture, 
this will become D5 branes wrapped on non-holomorphic cycles which can be deformed into holomorphic cycles in some limits. The extra deformation
necessary to obtain a supersymmetric configuration is related to an increase lifetime for these states.
 
One interesting direction would be to build supergravity solutions for D5 branes wrapped on $P^1$ cycles whose normal bundle components are not 
perpendicular, along the direction of \cite{bdt1}. This would allow us to study in more detail the geometrical transitions for these types of geometries.

\section{${\cal N} =1, SU(N_f) \times SU(N_c)$ Model}

Consider the ${\cal N} =2, SU(N_f) \times SU(N_c)$ model. This has two fields $\Phi, \tilde{\Phi}$ in the adjoint representation of $SU(N_f)$ and $SU(N_c)$
respectively. We consider the deformation  breaking the SUSY ${\cal N} =2 \rightarrow  {\cal N} =1$ as
\begin{equation}
\label{potelewoutlin}
W = \frac{1}{2} \tilde{\mu}  \tilde{\Phi}^2 +  \frac{1}{2} \mu  \Phi^2 +  \tilde{Q} (\lambda \Phi + \tilde{\lambda} \tilde{\Phi}) Q 
+ \xi \Phi +  \tilde{\xi} \tilde{\Phi}
\end{equation}
The F-term equations are
\begin{equation}
\label{1}
0 = \lambda \Phi Q + \tilde{\lambda} Q \tilde{\Phi},
\end{equation} 
\begin{equation}
\label{2}
0 = \lambda \tilde{Q} \Phi + \tilde{\lambda} \Phi \tilde{Q},
\end{equation}
\begin{equation}
\label{3}
0 = \mu \Phi + \lambda Q \tilde{Q} + \xi,~~\Phi =-\frac{\lambda Q \tilde{Q} + \xi}{\mu} 
\end{equation}
\begin{equation}
\label{4}
0 = \tilde{\mu} \tilde{\Phi} + \tilde{\lambda} \tilde{Q} Q + \tilde{\xi},~~\tilde{\Phi} =- \frac{\tilde{\lambda} Q \tilde{Q} + \xi}{\tilde{\mu}}
\end{equation}

If we consider that the mass of the field $\tilde{\Phi}$ is infinite and we integrate it out we get superpotential:
 \begin{equation}
\label{poteleintout}
W_{ele} = \frac{1}{2} \mu  \Phi^2 + \lambda \tilde{Q} \Phi Q  + \xi \Phi 
\end{equation}

The field $\Phi$ is a singlet of $SU(N_c)$ and in the adjoint representation of  $SU(N_f)$. The formula (\ref{poteleintout}) can be identified with the 
electric superpotential of \cite{giku1} as
\begin{equation}
\label{elident}
\Phi \leftrightarrow N,~~\mu \leftrightarrow - \alpha_e,~~\xi \leftrightarrow m_e,~~\lambda \leftrightarrow - \frac{1}{\Lambda}.
\end{equation}
or with the magnetic superpotential of  \cite{giku1} as
\begin{equation}
\label{magident}
\Phi \leftrightarrow M,~~\mu \leftrightarrow \alpha,~~\xi \leftrightarrow - m,~~\lambda \leftrightarrow \frac{1}{\Lambda}.
\end{equation}

By solving the F-term equations we see that in the  ${\cal N}=1$ theory there are quadratic terms
\begin{equation}
\frac{\lambda^2}{2 \mu} (\tilde{Q} Q)^2 +  \frac{\tilde{\lambda}^2}{\tilde{\mu}} (\tilde{Q} Q)^2
\end{equation}
and linear terms
\begin{equation}
\frac{\xi \lambda}{\mu} \tilde{Q} Q + \frac{\tilde{\xi} \tilde{\lambda}}{\tilde{\mu}} \tilde{Q} Q
\end{equation}
and the terms involving $\tilde{\mu}$ are zero in the limit $\tilde{\mu} \rightarrow 0$.

To get the ISS model \cite{iss}, we need to take extra limits in the electric theory  
\begin{equation}
\label{isslimit}
\mu \rightarrow \infty,~~\frac{\lambda^2}{\mu} \rightarrow 0,  \frac{\xi \lambda}{\mu} \rightarrow \mbox{finite} = m
\end{equation} 
where $m$ is exactly the mass of the electric quarks. The  ${\cal N} =2, SU(N_f) \times SU(N_c)$ theory is broken to an
${\cal N} =1, SU(N_f) \times SU(N_c)$ theory with light quarks if the quantity $ \frac{\xi \lambda}{\mu}$ is small. 

When $\frac{\lambda^2}{\mu}$ is also finite we get the extra term $\frac{\lambda^2}{\mu} (\tilde{Q} Q)^2$ and the 
${\cal N} =1$ superpotential is
\begin{equation}
\frac{\lambda^2}{2 \mu} (\tilde{Q} Q)^2 +  \frac{\xi \lambda}{\mu} \tilde{Q} Q
\end{equation}
The electric superpotential is
\begin{equation}
\frac{\lambda_e^2}{2 \mu_e} (\tilde{Q} Q)^2 +  \frac{\xi_e \lambda_e}{\mu_e} \tilde{Q} Q
\end{equation}
and the magnetic one is
\begin{equation}
\frac{\lambda_m^2}{2 \mu_m} (\tilde{Q} Q)^2 +  \frac{\xi_m \lambda_m}{\mu_m} \tilde{Q} Q
\end{equation}
A general SUSY solution is given by a vacuum expectation value for $\tilde{Q} Q$ given by
\begin{equation}
\label{vev}
\nu = \frac{\mu \xi}{\lambda}.
\end{equation}
Considering that the matrices in (\ref{vev}) have rank $k$, they will determine a breaking of the $SU(N_f) \times SU(N_c)$ group into
$SU(N_f-k) \times SU(N_c-k) \times SU(k)$. 

In order to obtain the results of \cite{giku1,giku2}, we consider the relations between the electric and magnetic variables
\begin{equation}
\lambda_{m} = - \lambda_e,~~\mu_{m} = - \frac{\lambda_{e}^2}{\mu_e},~~\xi_{m} = - \frac{\xi_{e}}{\lambda_{e} \mu_e}
\label{relation}
\end{equation}

\section{The corresponding Geometry}

As in \cite{tw2}, we start with a resolved ${\cal N} = 2, A_3$ singularity. Each $P^1$ cycle has a normal 
bundle:
\begin{equation}
X' = X,~~Y' = Y Z^2,~~Z'=1/Z
\end{equation}
The  ${\cal N} = 2, A_3$ singularity is then deformed into a collection of resolved conifold singularities, 
each one looking like
\begin{equation}
X' = X Z,~~Y' = Y Z,~~Z'=1/Z
\end{equation}

In the present work, we consider the deformation to geometries similar to the ones discussed in \cite{rtw}:
\begin{equation}
X' = X_r Z,~~Y' = Y Z,~~Z'=1/Z
\end{equation}
where the rotated direction $X_r$ is defined by
\begin{equation}
X_r = X - \frac{1}{m_{\mbox{adj}}} Y Z
\end{equation}
This direction is chosen such that in the blowdown map 
\begin{equation}
x~=~X~=X'Z',~y=ZY~=~Y',~u~=~ZX~=~X'~,~v~=~Y~=~Z'X'
\end{equation}
we find the deformation of the singular conifold
\begin{equation}
u~v~-~y~(x~-~\frac{1}{m_{\mbox{adj}}}~y) = 0.
\end{equation}

By rescaling $u$ and $v$, this geometry is T-dual to one with NS branes in the directions $y$ and $y~\mbox{cos}~\theta - x~\mbox{sin}~\theta$
where $\mbox{tan} \theta = m_{\mbox{adj}}$. In the electric theory we identify the direction $y$ with $v=x^4+ i x^5$ (the direction of unrotated 
NS branes), the direction $x$ with $w=x^8 + i x^9$ (the direction of NS branes rotated at 90 degrees angle) and denote
\begin{equation} 
v_{\theta}~=~w~\mbox{sin}~\theta~-~v~\mbox{cos}~\theta
\end{equation}
The electric theory contains from left to right an NS brane  in the direction $v_{\theta}$, one in the direction $v$ and one in the direction $w$.
The  ${\cal N} =1, SU(N_f) \times SU(N_c)$ model is obtained by putting $N_f$ D4 branes between $v_{\theta}$ and $v$ NS branes and 
$N_c$ D4 branes between the $v$ and $w$ NS branes. We chose this configuration because we do not want an extra adjoint in the gauge group 
(the reason for taking the color D4 branes between orthogonal NS branes) and want a finite mass adjoint field for the flavor group (the reason for taking 
the flavor D4 branes between non-orthogonal NS branes). 

To get the ISS model, in \cite{tw2} we considered the case $\theta = \pi_2$ in which case the branes $v_{\theta}$ and $w$ are parallel in the direction $w$. 
Their 
displacement in the $v$ direction is the mass of the quarks in the electric theory. In the geometrical engineering, the interval between 
the $v_{\theta}$ and $v$ NS branes becomes a $P^1$ cycle denoted $C_1$, the interval between the $v$  and $w$ NS branes becomes a $P^1$ cycle
 denoted $C_2$ and the 
interval between the $v_{\theta}$ and $w$ NS branes becomes a $P^1$ cycle denoted $C_3$.
 
The works \cite{beasley,hehana,vafafi} considered the Seiberg duality as a Toric duality. In our case, the 
Toric duality implies shrinking down the cycle $C_1$ and blowing up a cycle $C_4$.  
There is no problem to be encountered for massless quarks as the tachyon condensation occurs for branes on top of each other, but the 
situation is more involved for massive quarks. The electric geometry is Figure 1 and the magnetic theory in Figure 2. 

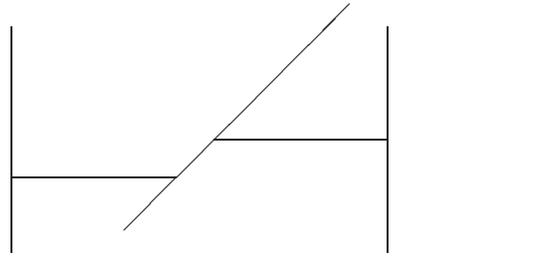
\begin{figure}
\begin{center}
\setlength{\unitlength}{1mm}
\begin{picture}(80,40)  
\put(5,15){\line(1,0){22}}
\put(32,20){\line(1,0){23}}
\put(5,5){\line(0,1){30}}
\put(75,5){\line(0,1){30}}
\put(20,8){\line(1,1){30}}
\put(55,5){\line(0,1){30}}
\end{picture}
\caption{Electric configuration of branes with $N_f$ Massive Quarks}
\end{center}
\end{figure}

\begin{figure}
\begin{center}
\setlength{\unitlength}{1mm}
\begin{picture}(80,40)
\put(32,20){\line(1,0){23}}
\put(27,15){\line(1,0){48}}
\put(5,5){\line(0,1){30}}
\put(75,5){\line(0,1){30}}
\put(20,8){\line(1,1){30}}
\put(55,5){\line(0,1){30}}
\end{picture}
\caption{Magnetic configuration of branes from $N_f$ Massive Quarks}
\end{center}
\end{figure}
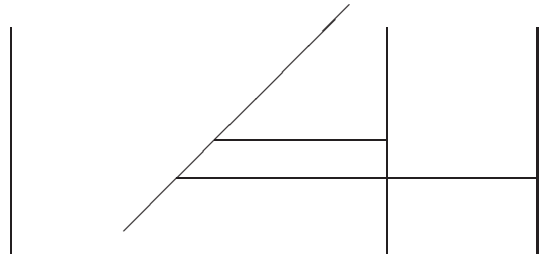

As discussed in \cite{tw2}, in the magnetic theory there is tachyon instability between $N_c$ anti D5 branes wrapped on the $C_2$ cycle and 
$N_f$ D5 branes wrapped on a $C_5=C_2+C_4$ cycle. Because the cycles are fixed by the rigidity of their normal bundles, 
the only way to have tachyon condensation is to deform the geometry in a non-holomorphic way. The end result 
is $N_f - N_c$ D5 branes wrapped on the $C_5=C_2+C_4$ cycle and $N_c$ D5 branes wrapped on a non-holomorphic cycle.

In the brane configuration picture, the main obstacle against a 
redistribution of D-branes on holomorphic cycles is the fact that the ends of D4 branes on the $v$ NS brane can touch each other
whereas the other ends cannot because they are stuck on parallel NS branes. In order to remove this problem, we need to 
make the two components of the normal bundle (or the NS branes) non-parallel. This is obtained by considering a 
field in the adjoint representation of the flavor group with finite mass.

The effect of the flop on the normal bundles is to exchange the directions $v$ and $w$. From the definition 
of $X_r$ we see that we also need to take $m_{\mbox{adj}} \leftrightarrow \frac{1}{m_{\mbox{adj}}}$. 
In terms of brane configurations the adjoint mass is $\mbox{tan}\theta$ where $\theta$ is the rotation angle and this 
implies that the relation between the rotation angles in the electric and magnetic pictures should be
\begin{equation}
\theta_{\mbox{electric}} = \pi/2 -  \theta_{\mbox{magnetic}}
\end{equation}

What terms do we have in the superpotential as a function of the rotation angles? In order to answer this, we need to review the appearance of the 
rotation angles in the adjoint masses and Yukawa couplings. The Yukawa coupling for  ${\cal N} =2, SU(N_f)$ with $N_c$ fundamental flavors $Q$ is
\begin{equation}
Q \Phi \tilde{Q}
\end{equation}
In the electric theory, 
this correspond to D4 branes between two parallel NS branes in the direction $v = x^4 + i x^5$. Consider now the rotation of one of the NS by an angle $\theta$
into direction 
\begin{equation}
v_{\theta}=~-~v~\mbox{cos}~\theta~+ w~\mbox{sin}~\theta
\end{equation}
The Yukawa coupling will acquire an extra term 
\begin{equation}
\lambda_e(\theta) = - \mbox{cos}~\theta
\end{equation}
The mass of the adjoint field is equal to $\mbox{tan}~\theta$ and this comes from the ratio of displacements in the $w$ and $v$ directions.  

In the magnetic theory, the Yukawa coupling for  ${\cal N} =2, SU(N_f)$ with $N_f - N_c$ fundamental flavors $q$ is
\begin{equation}
\label{magneticp}
q \Phi \tilde{q}
\end{equation}
This corresponds to D4 branes between two parallel NS branes in the direction $w = x^8 + i x^9$. The coefficient is 1 in (\ref{magneticp}) if we start with
orthogonal NS branes in the electric theory and just keep track of the exchange $v \leftrightarrow w$ when considering the flop.

The situation becomes more complicated if some NS branes are rotated in the electric picture. In this case we need to keep track of the different normal bundles to 
various $P^1$ cycles wrapped with D5 branes. We need to make the following identifications:

1) unrotated directions $v$ and $w$ are parts of the normal bundle to the $P^1$ and they are exchanged during the flop. 

2) the angles of rotation in the electric theory are with respect to the axis $v$. If the axis $v$ and $w$ are exchanged, the rotation angle $\theta$ with 
respect to the axis $v$ becomes rotation angle $\theta$ with respect to the axis $w$.This is equivalent to having the exchange between  
the rotation angle $\theta$ with respect to the axis $v$ and the rotation angle $\pi/2 - \theta$ with respect to the axis $v$.

The deformation to geometries are now
\begin{equation}
X' = X Z,~~Y' = Y_r Z,~~Z'=1/Z
\end{equation}
where the rotated direction $X_r$ is defined by
\begin{equation}
Y_r = Y - \frac{1}{m_{\mbox{adj}}} X Z
\end{equation}
In the blowdown map 
\begin{equation}
x~=~X~=X'Z',~y=ZY~=~Y',~u~=~ZX~=~X'~,~v~=~Y~=~Z'X'
\end{equation}
we find the deformation of the singular conifold
\begin{equation}
u~v~-~x~(y~-~\frac{1}{m_{\mbox{adj}}}~x) = 0.
\end{equation}
By identifying as before $x$ with $w$ and $y$ with $v$ we see that the NS branes are now one in the direction $w$ and the other in the direction
\begin{equation}
v_{\theta'}=  v~\mbox{sin}~\theta'~-~w~\mbox{cos}~\theta'
\end{equation}
where $\theta'=\pi/2 - \theta$.

Now, to compare with the results of \cite{giku2}, we need to consider a different sign for the masses of the ${\cal N} =2$ adjoint fields in the electric and 
magnetic theories. This would imply that we take $\theta_m = - \theta_e$ such that 
\begin{equation}
\mbox{tan} \theta_m = - \mbox{tan} \theta_e
\end{equation}
which then gives
\begin{equation}
\mbox{cos}~\theta'_m = \mbox{sin} \theta_m = - \mbox{sin} \theta_e =   - \mbox{cos} \theta'_e
\end{equation}
which states  that the term multiplying $w$ has a positive sign.

 This implies the magnetic Yukawa coupling being proportional to
\begin{equation} 
\lambda_m(\theta) = \mbox{cos} \theta'
\end{equation}
and the mass of the magnetic adjoint field being proportional to  $\mbox{tan} \theta$ is the electric one was  $- \mbox{tan} \theta$.
If we care about the dimensions, the Yukawa couplings should contain a $\frac{1}{\Lambda}$ factor.

Therefore the sign of the Yukawa couplings and the mass of the adjoint fields have a different sign in the magnetic and electric theories. 
In order to reach the results of \cite{giku1},\cite{giku2} we chose the convention that the magnetic terms are positive and the electric terms are negative.

What about the annihilation between the branes and antibranes? In the magnetic picture, the cycle wrapped by the $N_f$ D5 branes  can slide. 
The annihilation begins on the $v$ line and propagates to the right. If the $N_f$ D5 branes were located at $(v_2,0)$ in their normal bundle and wrapped on the cycle
$C_5=C_2 + C_4$,  the part lying on $C_2$ annihilates the $N_c$ anti D4 branes to give $N_f - N_c$ D5 branes on the $C_2$ cycle 
and the  part is now on a $C_4$ NS brane touching its normal bundle at a point $(0,-w_2 \mbox{cot} \theta)$. 
This is the dual $SU(N_f - N_c)$ theory with $N_f$ fundamental flavors. 

\subsection{Complex Deformations and Rearrangement of Cycles}

As discussed in detail in \cite{tw2}, the deformation of ${\cal N} =2$ geometries give rise to  ${\cal N} =1$ geometries with 
normalizable and non-normalizable deformations. The normalizable deformations are the ones which can go through a geometric transition 
to study strong coupling effects  in the field theory on the wrapped D5 branes. The non-normalizable
deformations are the ones corresponding to masses or vevs for the bifundamental ${\cal N} =2$ quarks.

By starting with a resolved $A_n$ singularity, adding $N_i$ D5 branes on each $i$ cycle, we get the gauge group
\begin{equation}
\prod_{i=1}^{n} SU(N_i).
\end{equation}
By adding quadratic terms in the adjoint fields, we get an  ${\cal N} =1$ geometry with 
\begin{equation}
n(n+1)/2~~~~~\mbox{normalizable deformation}
\end{equation}
and 
\begin{equation}
n(n-1)/2~~~~~\mbox{non-normalizable deformation}
\end{equation}
For a deformation of $A_3$, we have 6 normalizable and 3 non-normalizable deformations. 
In \cite{tw2} we saw that the mismatch of the non-normalizable deformations could be used as an explanation why the ISS solution cannot 
be constructed with D5 branes wrapped on holomorphic cycles. This was due to the fact that in the electric picture one has 
two masses for the adjoint fields whereas in the magnetic picture one has mass only for one adjoint fields. The tachyon condensation on the 
magnetic theory corresponded to non-holomorphic deformations of the cycles.

For the present discussion we have two masses for the adjoint fields in both electric and magnetic picture so there is no mismatch. 
The non-normalizable cycles are visible in both electric and magnetic pictures:

$\bullet$ in the electric picture the non-normalizable deformations corresponds to the masses for the light quarks. In brane configurations, the
non-normalizable deformation corresponds to displacing the $N_f$ D4 branes from the point $(0,0)$ to a point $(v,0),~v \ne 0$. 

$\bullet$ in the magnetic picture the non-normalizable deformations corresponds to the vevs for the adjoint field of the flavor group.
In brane configuration this correspond to displacing the flavor  $N_f$ D4 branes from the point $(0,0)$ to the point $(0,w),~w \ne 0$.

$\bullet$ the tachyon condensation correspond to exchanging the two types of non-normalizable deformations. The displacement in the $v$ 
direction between the flavor and color branes is mapped into a displacement in the $w$ direction.

$\bullet$ from geometry we see that by starting with an electric displacement to $(v,0)$, considering the flop and then the tachyon 
condensation by sliding the $N_f$ branes along the $v_{\theta}$ direction, we get a displacement to $(0,v~\mbox{cot}~\theta_m)$.
Remembering that $\theta_m = - \theta_e$, it results that the displacement is  to $(0,- v~\mbox{cot}~\theta_e)$

\subsection{ ${\cal N} =2, SU(N_f) \times SU(N_c) \rightarrow  {\cal N} = 1, SU(N_f-k) \times SU(N_c- k) \times SU(k)$}
What happens if some of the electric quarks have a finite vacuum expectation value? As discussed in the introduction, the relation between the 
masses of the adjoints and the vev for the electric quarks is
\begin{equation}
\label{value}
<Q \tilde{Q}> = \frac{\mu}{\frac{1}{m_{N_f}} +  \frac{1}{m_{N_c}}}
\end{equation}
If $m_{N_f}$ is finite and we have a rotated NS brane in the $v_{\theta}$ direction, 
there are some extra solutions, as pointed out in \cite{giku1,giku2}. In terms of D4 branes, they are obtained by displacing $k$ D4 branes in the 
$w$ direction \footnote{From the  ${\cal N} =2$ Yukawa coupling, the displacement along the $v$ direction is always associated to the mass
of the electric quarks.}. In terms of D5 branes, the extra wrapped D5 branes appear because for the deformed $A_2$ singularity with 
quadratic superpotentials there are 3 normalizable deformation which correspond to $P^1$ cycles and we can wrap 
$N_f - k$ D5 branes , $N_c-k$ D5 branes and $k$ D5 branes on each of them respectively. The color $P^1$ cycle is wrapped by $N_c-k$ D5 branes and 
the flavor $P^1$ cycles by $N-f - k$ and $k$ D5 branes respectively.

As discussed in \cite{dot1}, for an $A_2$ singularity deformed by a superpotential 
\begin{equation}
W_1(v) + W_2(v),~~ v =~\mbox{vev of}~\Phi_1~\mbox{or}~\Phi_2
\end{equation}
the position of the D4 branes is given by 
\begin{equation}
\label{position}
W_1'(v)=0~~~~W_2'(v)=0~~~\mbox{and}~~~~W_1'(v) + W_2'(v)=0
\end{equation}
when one of the NS branes is unrotated. It is more convenient to consider that the unrotated NS brane is the $v_{\theta}$ brane such that the 
$v$ NS brane becomes rotated by an angle $- \theta$ and the $w$ NS brane becomes rotated by an angle $\pi/2 - \theta$. The equation
(\ref{position}) becomes
\begin{equation}
\label{positionthetaw}
W_1'(v_{\theta})=0~~~~~W_2'(v_{\theta})=0~~~\mbox{and}~~~~~W_1'(v_{\theta}) + W_2'(v_{\theta})=0
\end{equation}  
The D4 branes located at $W_1'(v_{\theta})=0$ are between the $v_{\theta}$ NS brane and the $v$ NS brane, the D4 branes located at 
$W_2'(v_{\theta})=0$ are between the $v$ NS brane and the $w$ NS' brane and the D4 branes located at $W_1'(v_{\theta}) + W_2'(v_{\theta})=0$ 
are between the $v_{\theta}$ NS brane and the $w$ NS' brane.

We consider now a flop of the color $P^1$ cycle which means that the normal bundles to the other $P^1$ cycles are also changed. 

\begin{figure}
\begin{center}
\setlength{\unitlength}{1mm}
\begin{picture}(80,40)  
\put(5,15){\line(1,0){22}}
\put(32,20){\line(1,0){23}}
\put(0,5){\line(1,2){17}}
\put(75,5){\line(0,1){30}}
\put(20,8){\line(1,1){30}}
\put(55,5){\line(0,1){30}}
\end{picture}
\caption{$k=0$ Electric configuration with $v_{\theta'}$ NS brane}
\end{center}
\end{figure}
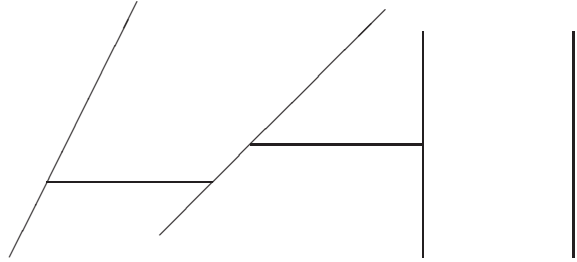

\begin{figure}
\begin{center}
\setlength{\unitlength}{1mm}
\begin{picture}(80,40)
\put(32,20){\line(1,0){23}}
\put(27,15){\line(1,0){27}}
\put(56,15){\line(1,0){24}}
\put(5,5){\line(0,1){30}}
\put(75,5){\line(1,2){17}}
\put(20,8){\line(1,1){30}}
\put(55,5){\line(0,1){30}}
\end{picture}
\caption{$k=0$ Magnetic before tachyon condensation}
\end{center}
\end{figure}
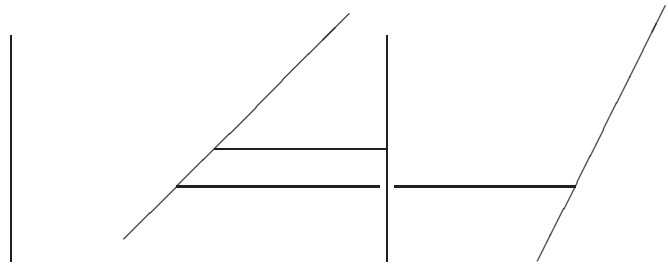

\begin{figure}
\begin{center}
\setlength{\unitlength}{1mm}
\begin{picture}(80,40)
\put(32,20){\line(1,0){23}}
\put(55,15){\line(1,0){25}}
\put(5,5){\line(0,1){30}}
\put(75,5){\line(1,2){17}}
\put(20,8){\line(1,1){30}}
\put(55,5){\line(0,1){30}}
\end{picture}
\caption{$k=0$ Magnetic after tachyon condensation}
\end{center}
\end{figure}
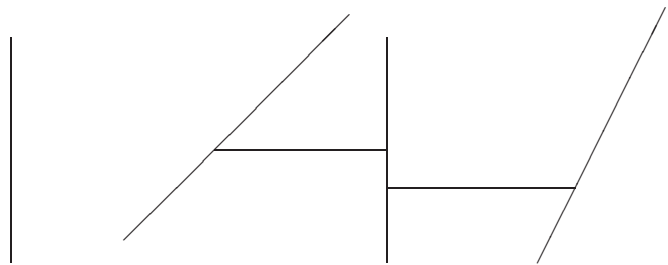

\begin{figure}
\begin{center}
\setlength{\unitlength}{1mm}
\begin{picture}(80,40)  
\put(5,15){\line(1,0){22}}
\put(32,20){\line(1,0){23}}
\put(12,30){\line(1,0){28}}
\put(44,30){\line(1,0){11}}
\put(0,5){\line(1,2){20}}
\put(75,5){\line(0,1){30}}
\put(20,8){\line(1,1){30}}
\put(55,5){\line(0,1){30}}
\end{picture}
\caption{$k \ne 0$ Electric configuration}
\end{center}
\end{figure}
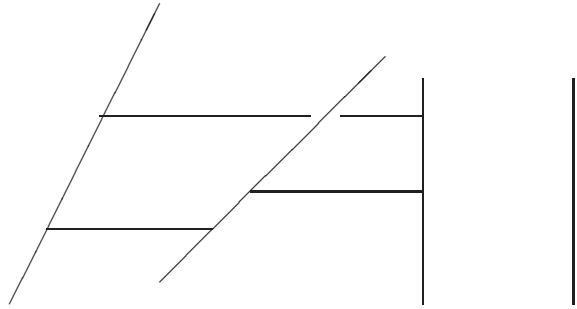

\begin{figure}
\begin{center}
\setlength{\unitlength}{1mm}
\begin{picture}(80,40)
\put(32,20){\line(1,0){23}}
\put(27,15){\line(1,0){27}}
\put(56,15){\line(1,0){24}}
\put(55,30){\line(1,0){32}}
\put(5,5){\line(0,1){30}}
\put(75,5){\line(1,2){17}}
\put(20,8){\line(1,1){30}}
\put(55,5){\line(0,1){30}}
\end{picture}
\caption{$k \ne 0$ Magnetic before tachyon condensation}
\end{center}
\end{figure}

\begin{figure}
\begin{center}
\setlength{\unitlength}{1mm}
\begin{picture}(80,40)
\put(32,20){\line(1,0){23}}
\put(55,15){\line(1,0){25}}
\put(5,5){\line(0,1){30}}
\put(75,5){\line(1,2){17}}
\put(20,8){\line(1,1){30}}
\put(55,5){\line(0,1){30}}
\put(56,30){\line(1,0){31}}
\put(42,30){\line(1,0){12}}
\end{picture}
\caption{$k \ne 0$ Magnetic after tachyon condensation}
\end{center}
\end{figure}

The reference NS brane is now on the $v_{\theta'}$ direction and the positions of the D4 branes are given by
\begin{equation}
\label{positionthetav}
W_1'(v_{\theta'})=0~~~~~W_2'(v_{\theta'})=0~~~\mbox{and}~~~~~W_1'(v_{\theta'}) + W_2'(v_{\theta'})=0
\end{equation} 
The middle NS brane is considered to be rotated from the direction $v_{\theta'}$ into direction $w$ and the other NS brane is considered to be 
rotated from the direction $v_{\theta'}$ into direction $v$. The D4 branes located at $W_1'(v_{\theta'})=0$ are 
between the $v_{\theta'}$ NS brane and the $w$ NS brane, the D4 branes located at 
$W_2'(v_{\theta'})=0$ are between the $w$ NS brane and the $v$ NS' brane and the D4 branes located at $W_1'(v_{\theta'}) + W_2'(v_{\theta'})=0$ 
are between the $v_{\theta'}$ NS brane and the $v$ NS' brane.

Now, if we consider the direct mapping of the cycles and the D4 branes during the flop, the conclusion appears to be different. The 
D4 branes appear to be extended between the same NS branes as in the electric geometric. But this would make sense only if the directions
$v_{\theta}$ and $v_{\theta'}$ were identical which they are not. This is the reason why only the $W_2'=0$ D4 branes are located between the same 
NS branes as in the electric case whereas the other two stacks change their ending points. 

In terms of the geometrical picture, we now discuss the change in the cycles wrapped by the D5 branes. By denoting the cycles in the same manner as in the 
previous subsection, we have the following wrapped branes:

$\bullet$ in the electric theory the $N_f - k$ D5 branes wrap the $C_1$ cycle, the $N_c - k$ D5 branes wrap the cycle $C_2$ and the 
$k$ D5 branes wrap the cycle $C_3$. The cycle $C_1$ has the normal bundle $v_{\theta}, v$, the cycle $C_2$ has normal bundle 
$v,w$ and the cycle $C_3$ has normal bundle $v_{\theta}, w$. 

$\bullet$ in the magnetic theory obtained after the flop, there are $N_c - k$ anti D5 branes wrapping the cycle $C_2$, $k$ D5 branes wrapping the 
$C_4$ cycle and $N_f-k$ D5 branes wrapping the $C_5$ cycle. The cycle $C_4$ has normal bundle $w,v_{\theta'}$, the cycle $C_2$ has normal bundle 
$v,w$ same as before and the cycle $C_5$ has normal bundle $v, v_{\theta'}$.

$\bullet$ if we only look at the magnetic theory, this would involve the breaking of the symmetry 
\begin{equation}
SU(N_f) \times SU(N_f-N_c) \rightarrow  SU(N_f-k) \times SU(N_f-_Nc-k) \times SU(k)
\end{equation}
This would involve $N_f-N_c-k$ D5 branes wrapped on the $C_2$ cycle, $N_f-k$ D5 branes wrapped on the $C_5$ cycle and $k$ branes wrapped on the $C_4$
cycle.
    
Because they both represent the end of a Seiberg duality, the last two cases should be identical. How do we obtain one from the other? 
This is achieve by dynamical process due to a tachyon condensation which we discuss next.  

After the flop, the cycle with wrapped color branes changes the orientation and the wrapped $N_c - k$ D5 branes become $N_c - k$ wrapped anti D5 branes. 
There are now two types of tachyons between D5 branes and anti D5 branes:

1) tachyon between the  $N_c-k$ anti D5 branes and the $N_f - k$ D5 branes.

2)  tachyon between the  $N_c-k$ anti D5 branes and the $k$ D5 branes.

Due to the condensation of the tachyon, the geometry is dynamically modified. In principle we would have two possibilities \footnote{We remember that 
there is a relation $\theta_{m} = \pi/2 - \theta_{e}$ between the angles of rotation in the magnetic and electric theories. This means that a small
angle in the electric theory means a large angle in the electric theory. Here we refer to the adjoint fields in the magnetic theory i.e. to the 
rotation angles in the magnetic theory.} :

1) the magnetic angle of rotation i.e. the mass of the magnetic adjoint field is small. In this case the value of the expectation value in (\ref{value}) is
smaller than the mass of the flavors
\begin{equation} 
<Q \tilde{Q}> = \frac{\mu}{\frac{1}{m_{N_f}}} < \mu
\end{equation}
In this case the displacement of the $k$ D5 branes along the $w$ direction with respect to the $N_c-k$ anti D5 branes is smaller than the 
displacement of the $N_f - k$ D5 branes with respect to the  $N_c-k$ anti D5 branes.  Therefore the tachyon condensation would occur 
first between the $k$ D5 branes and  the $N_c-k$ anti D5 branes to give  
\begin{equation}
N_c-2k~~\mbox{anti D5 branes}~+~  
k~~\mbox{D5 branes between}~~v, v_{\theta'}
\end{equation}
There are still tachyons between the $N_c-2k$ anti D5 branes and the $N_f-k$ D5 branes and the second tachyon condensation will give as in the previous 
subsection
\begin{equation}
N_f-N_c-k~~\mbox{color D5 branes}~+~  
N_f-k~~\mbox{D5 branes between}~~w, v_{\theta'}
\end{equation}

2) the magnetic angle of rotation i.e. the mass of the magnetic adjoint field is big. In this case the value of the expectation value in (\ref{value}) is
larger than the mass of the flavors
\begin{equation} 
<Q \tilde{Q}> = \frac{\mu}{\frac{1}{m_{N_f}}} >> \mu
\end{equation}
In this case the displacement of the $k$ D5 branes along the $w$ direction with respect to the $N_c-k$ anti D5 branes is larger than the 
displacement of the $N_f - k$ D5 branes with respect to the  $N_c-k$ anti D5 branes. 
The tachyon condensation occurs first between the  $N_c-k$ anti D5 branes and the $N_f - k$ D5 branes to give $N_f - N_c$ color D5 branes 
and $N_f-k$ flavor D5 branes between $w$ and $v_{\theta'}$, as in previous subsection.

Which way to go? To answer this question, we remember the results of \cite{mukhi} whose conclusion of their paper was that, when D4 and anti D4 branes end on the
opposite sides of an NS branes, they repel each other. Their argument uses the fact that the end of the D4 branes are charged 3-branes in the 
NS world volume. If one dimensionally reduces over these 3 dimensions, the end are vortices living on the reduces NS world volume.
We can apply this argument to the two types of pairs of D-branes anti D-branes:

a) $k$ D4 branes and $N_c-k$ anti D4 branes ending on an NS brane at opposite sides.  The vortices carry the same charge under the gauge 
field and they repel. Therefore we cannot have this as the first tachyon condensation.

b) $N_f-k$ D4 branes and $N_c-k$ anti D4 branes ending on an NS brane at the same side. In this case the vortices carry different charges 
under the gauge field and they attract. This is then the first tachyon condensation. 
The tachyon condensation changes the cycle wrapped by the 
$N_f-k$ D5 branes which will now have the normal bundle $w, v_{\theta'}$. 

What is the geometrical implication? We saw before that there are  $N_f-k$ D5 branes on the $C_5$ and $N_c-k$ anti D5 branes on the $C_2$ cycle. 
This configuration is unstable and there is a tunnelling process taking us to a stable configuration obtained by splitting the $C_5$ cycle into 
$C_2$ and $C_4$ such that there are now $N_f-k$ D5 branes on both $C_2$ and $C_4$ and then annihilating the  $N_f-k$ D5 branes on $C_2$
with the  $N_c-k$ anti D5 branes. The result is  $N_f-k$ D5 branes on $C_4$ and  $N_f-N_c$ D5 branes on both $C_2$.

This is not the end of the story because we did not specify what happens to the $k$ D5 branes wrapped on the cycle $C_4$. As the cycle $C_5$ changes its 
normal bundle, the cycle $C_4$ also needs to change its normal bundle. This could be understood in the following way:

- as discussed before, the separation between the cycles is related to the expectation values of the various fields in the theory. These fields can either be
the gauge singlet $\Phi$ or the gauge invariant combination between the dual quarks $q \tilde{q}$. 

- the vev for the gauge singlet $\Phi$ or for $q \tilde{q}$ should be measured along different directions in the geometry.
Before tachyon condensation, we consider that the vev for $\Phi$ should be measured along the direction $v$ and the vevs for    $q \tilde{q}$ should be measured 
along the direction $w$. 

- after tachyon condensation, if the   $N_f-k$ D5 branes on the $C_5$ cycle change into  $N_f-k$ D5 branes on the $C_4$ cycle, the vev for $\Phi$ should now be measured 
along the direction $w$ which means that the vevs for  $q \tilde{q}$ should now be measured along the direction $v$. This implies that the 
displacement of the   $k$ D5 branes with respect to the color branes should be on the direction $v$. The only way to do this and have the 
$k$ D5 branes on a holomorphic cycle is to put them on the cycle $C_5$. 

- the process implies that, at the first instance,the $k$ D5 branes become wrapped on a cycle which is non-holomorphic and the energy of the $k$ D5 branes 
increases after the deformation.
 We can use the the same vortex argument to state that the $k$ D5 branes and the $N_f - N_c$ color D5 branes attract each other.The system is unstable 
and the excess energy is lost  when $k$ of the  $N_f - N_c$ color D5 branes combine with the $k$ D5 branes to give $k$ D5 branes on the cycle $C_5$. 
We leave the detailed description of the combined process of tachyon condensation and cycle redistribution for future work.

The solutions obtained in this section are supersymmetric because the relation  (\ref{value}) tells us that there is a compatible deformation 
\begin{equation}
C_4 \leftrightarrow C_5,~~ C_5 \leftrightarrow C_4
\end{equation}
In the next section we discuss the non-supersymmetric solutions where there are flavor branes wrapped on cycles $C_5' \ne C_5$
which are not exchangeable with $C_4$ and remain non-holomorphic after tachyon condensation.

\section{Metastable Vacua}

Beside the SUSY vacua, the authors of \cite{giku1,giku2} have also discussed a large set of metastable non-SUSY vacua. 
They appear when we have a further breaking of the flavor group 
\begin{equation}
SU(N_f-k) \rightarrow SU(N_f - k - n) \times SU(n)
\end{equation}
such that the masses of the $SU(N_f - k - n)$ flavors satisfy  (\ref{value}) whereas the masses of the $SU(n)$ flavors do not satisfy (\ref{value}).
This will imply that there are D4 branes whose positions are not described by (\ref{positionthetav}). 

In the electric theory this means that we have $N_f-k-n$ D5 branes on a $C_1$ cycle, $n$ D5 branes on a $C_1'$ cycle, $N_c$ D5 branes on a 
$C_2$ cycle and $k$ D5 branes on a $C_3$ cycle.  The cycles $C_1$ and $C_1'$ are displaced on the $v$ direction. 

What happens during the Seiberg dualities? The result of the flop is 
$N_c - k$ anti D5 branes on $C_2$, $N_f - k - n$ D5 branes on a $C_5$ cycle , $n$ D5 branes on a $C_5'$ cycle and $k$ D5 branes on a $C_4$ cycle.
The cycles $C_5$ and $C_5'$ are also displaced in the   $v$ direction. 

How does the tachyon condensation occur? As discussed in \cite{giku1}, the validity 
of their analysis holds in the limit that the coefficients in (\ref{potelewoutlin}) obey
\begin{equation}
\mu << \xi
\end{equation}
which implies a very small magnetic angle of rotation. We can again
use the results of \cite{mukhi} to argue that the  $N_c - k$ anti D5 branes and the $k$ D5 branes repel each other and no tachyon condensation takes 
place between them.

The tachyon condensation occurs first between the  $N_c - k$ anti D5 branes and the 
$N_f-k-n$ D5 branes. The result is $N_f - N_c-n$ color D5 branes. 

The $N_f - k - n$ D5 branes change the wrapped cycle from $C_5$ to $C_4$. What about the stack of $n$ D5 branes and the stack of $k$ D5 branes?
Because of the $C_4 \leftrightarrow C_5$ interchange, the stack of $k$ D5 branes becomes wrapped on $C_5$ due to the existence of the relation (\ref{value}).
As the position of the   $n$ D5 branes does not satisfy  (\ref{value}), they do not become wrapped on some $C_4'$ cycle. The cycle they wrap, denoted by 
$\tilde{C}$, is a non-holomorphic deformation of the original $C_5$ cycle and this makes the configuration non-supersymmetric. 

The non-supersymmetric configuration can decay into supersymmetric ones. As a function of the location of the $n$ D5 branes we have 2 possibilities 
for such supersymmetric configurations:

1) the  $n$ D5 branes are close to the $k$ D5 branes. The switch $C_4 \leftrightarrow C_5$ is performed at the same time as $\tilde{C}$ changes into a
holomorphic cycle and  we get in the final configuration $k+n$ D5 branes wrapped on the same $C_5$ cycle. The metastability is related to the 
deformation of the non-holomorphic cycle $\tilde{C}$ into $C_5$. It would be interesting to show that the duration of this deformation is larger that the
time necessary for the tachyon to condense, a crucial requirement for metastability.  
 
The final configuration is the same as the SUSY configuration obtained as
\begin{equation}
{\cal N} =2, SU(N_f) \times SU(N_c) \rightarrow  {\cal N} = 1, SU(N_f-k-n) \times SU(N_c- k-n) \times SU(k+n)
\end{equation}
with the emphasis that there are some intermediate steps which should be covered in a longer time than the usual 
\begin{equation}
{\cal N} =2, SU(N_f) \times SU(N_c) \rightarrow  {\cal N} = 1, SU(N_f-k) \times SU(N_c- k) \times SU(k)
\end{equation}
breaking.

2) the $n$ D5 branes are very far from the $k$ D5 branes and the color D5 branes and are close to the 
$N_f-k-n$ D5 branes.  The switch $C_4 \leftrightarrow C_5$ is again performed while $\tilde{C}$ changes into a
holomorphic cycle and we get in the final picture a number of 
\begin{equation}
N_f - k - n + n = N_f - k~~\mbox{D5 branes wrapped on}~~~C_4~~\mbox{cycle}.
\end{equation}
The metastability is now related to the deformation of the non-holomorphic cycle $\tilde{C}$ into $C_4$. The final configuration is 
identical to the one where we start from an unbroken $SU(N_f-k)$ group in the electric theory.
Also for this case it would interesting to understand why the deformation of the cycle should take much longer time than the 
tachyon condensations.

The above discussion holds for the range $1 \le n \le N_f-N_c-k$, when the number $N_f - N_c - n - k$ of color D5 branes is positive.
What happens when $n >  N_f-N_c-k$ i.e.  $N_f - N_c - n - k < 0$? In this case the first type of tachyon condensation does not occur so the solution is more 
stable. All the other cases considered in \cite{giku1,giku2} can be expressed in terms of tachyon condensations between pairs of branes and anti branes and
rearranging of cycles to holomorphic embeddings.

\section*{Acknowledgements}

The work of Radu Tatar is funded by PPARC. Radu Tatar would like to thank the organisers of the program Strong fields,Integrability and
Strings at the Isaac Newton Institute for Mathematical Sciences during which this work was initiated.

\end{document}